\documentclass{tlp}

\usepackage{amsmath}
\usepackage{graphicx}
\usepackage{multirow}
\usepackage{amssymb}
\usepackage{xcolor}
\usepackage{url}
\usepackage{textcomp}
\newtheorem{example}{Example}

\begin{document}

\pagerange{\pageref{firstpage}--\pageref{lastpage}}

\title[Tau Prolog: A Prolog interpreter for the Web]{Tau Prolog: A Prolog interpreter for the Web \thanks{This work has been partially supported by the State Research Agency (AEI) of the Spanish Ministry of Science and Innovation under grant PID2019-104735RB-C42 (SAFER).}}

\author[José A. Riaza]
	{{JOSÉ A. RIAZA}\\
    Department of Computing Systems, University of Castilla-La Mancha, \\ 02071 Albacete (Spain)\\
	\email{JoseAntonio.Riaza@uclm.es}
	}

\maketitle

\begin{abstract}
Tau Prolog is a client-side Prolog interpreter fully implemented in JavaScript, which aims at implementing \textcolor{black}{the} ISO Prolog Standard. Tau Prolog has been developed to be used with either Node.js or a browser seamlessly, and therefore, it has been developed following a non-blocking, callback-based approach to avoid blocking web browsers. Taking the best from JavaScript and Prolog, Tau Prolog allows the programmer to handle browser events and manipulate the Document Object Model (DOM) of a web using Prolog predicates. In this paper we describe the architecture of Tau Prolog and its main packages for interacting with the Web, and we present its programming environment. 
\end{abstract}

\begin{keywords}
Tau Prolog, Logic Programming, Prolog Interpreter, JavaScript
\end{keywords}

\section{Introduction}

HTML, CSS and JavaScript are the basic building blocks of the Web. JavaScript is a programming language that allows the programmer to implement complex things on web pages. The standard for JavaScript is ECMAScript \cite{ECMA402,ECMA262}. A web browser provides an ECMAScript host environment for client-side computation,
which in turn provides a means to attach scripting code to events such as change of focus, form submission, and mouse actions. The scripting code is reactive to user interaction, and there is no need for a main program.
JavaScript has a concurrency model based on an event loop, which is responsible for executing the code, collecting and processing events, and executing queued sub-tasks. A JavaScript runtime is single threaded and uses a message queue, where each message is completely processed before any other message is treated.\footnote{See \cite{mdn2021concurrency} for more details about the event loop.}
A downside of this model is that if a message takes too long to complete, the web application is unable to process user interactions.
%
%
%

While most online Prolog systems, such as SWISH introduced by \cite{WielemakerLR15}, are remote servers with an installed version of the Prolog system which receive code, execute it and return the results, Tau Prolog \cite{riaza2021tau} is fully implemented on JavaScript, and therefore the code can be analyzed and executed directly on the client-side. This allows Tau Prolog to easily create interfaces for the JavaScript Web APIs in order to interact with web pages using Prolog. There are other approaches, such as Prolog to JavaScript compilers \cite{morales2012lightweight}, or systems like Yield Prolog \cite{yield2022} that allow to simulate the behavior of logic programming in JavaScript through generators. SWI-Prolog pengines \cite{lager_wielemaker_2014} also provides an abstraction to create and query Prolog engines over HTTP from JavaScript running in a web client.
{\color{black} In the last years, some Prolog systems such as Ciao Prolog\footnote{\url{https://github.com/ciao-lang/ciaowasm}}, SWI-Prolog\footnote{\url{https://www.swi-prolog.org/build/WebAssembly.html}} and Trealla Prolog\footnote{\url{https://github.com/trealla-prolog/trealla}} added support for WebAssembly compilation  \cite{garcia22} which allows them to run in most major browsers.} 
In addition, we found a few JavaScript Prolog implementations like Tau Prolog really capable of parsing and executing Prolog code in a browser, such as \textcolor{black}{jsProlog\footnote{\url{https://github.com/cubiwan/jsProlog}}, Dogelog\footnote{\url{https://www.xlog.ch/izytab/doclet/en/docs/23_products/10_jekrun/package.html}} and hitchhiker Prolog\footnote{\url{https://github.com/CapelliC/hhprolog}} --an implementation of a Prolog virtual machine proposed by \cite{tarau2018}--}. However, Tau Prolog provides a full client-side Prolog implementation including features like modules, meta-predicates mechanism, DCGs, term and goal expansion, and ISO Prolog predicates, that are not present in these \textcolor{black}{JavaScript} implementations. Furthermore, Tau Prolog is based on a simple and extensible representation, in order to be easy to modify at runtime by users (something difficult to achieve with \textcolor{black}{approaches such as pengines or WebAssembly}).

With the aim of mitigating the negative effects of the JavaScript concurrency model, Tau Prolog has been developed following a non-blocking, callback-based approach, since looking for computed answers in Prolog can be a heavy task for certain programs.
Web workers \cite{mdn2022workers} provide another means for web to execute scripts in background threads without interfering with the user interface, but they don't have direct access to the DOM, it is necessary to execute an entire script, and only serialized objects can be exchanged.
For these reasons, the callback-based approach was chosen. Furthermore, users can embed a Tau Prolog computation into a web worker using the current interface.

The source code of Tau Prolog is available on GitHub,\footnote{\url{https://github.com/tau-prolog/tau-prolog}} and it is freely distributed with npm.\footnote{\url{https://www.npmjs.com/package/tau-prolog}}
Tau Prolog aims at implementing the ISO Prolog Standard \cite{ISO13211}, designed to promote the applicability and portability of Prolog text and data among several data processing systems. The main purpose of Tau Prolog is to provide a complete Prolog system that can be embedded in a web page, for use in web applications that can benefit from Prolog, such as in the validation and intelligent completion of web forms, and in tasks that require dynamic adaptions of the DOM itself based on logical rules. In last years, Tau Prolog has been gaining popularity, both in academia
\cite{brodo2021logical,kirchev2019demonstrating,latypova2020decision} and in real world applications \cite[Logtalk]{yarnpkg,TheLogtalkHandbook}, showing some of the potential uses of Tau Prolog. Furthermore, the Tau Prolog sandbox provides an ideal environment for teaching, since it does not require any installation and allows derivations to be graphically visualized. Moreover, Tau Prolog is a good option for teaching thanks to its commitment to the Prolog ISO standard and its free accessibility. \textcolor{black}{In addition, it keeps code and data (such as family relations of students, which are frequently used in Prolog exercises) confidential on the client.}


In this paper we describe the architecture of Tau Prolog and its main packages for interacting with the Web.
The structure of this paper is as follows.
In Section \ref{sec:architecture} we describe the Tau Prolog architecture for loading programs and querying goals asynchronously. Then Section \ref{sec:web-interaction} is devoted to describe the main packages of Tau Prolog for the web interaction. In Section \ref{sec:tools} we show the programming environment for Tau Prolog. Finally, in Section \ref{sec:benchmarks} we present some benchmarks measuring the performance of Tau Prolog.

\section{Tau Prolog architecture}
\label{sec:architecture}

All the Tau Prolog functionality is embedded in a JavaScript object named \texttt{pl}, which is visible in the global scope after importing the library. In this section, we introduce the main interfaces to work asynchronously with Tau Prolog. 

\subsection{Prototypes and Prolog objects}
\label{sec:objects}

Here, we use a ground representation in order to store and manipulate Prolog terms with JavaScript.
JavaScript is a prototype-based language \cite{mdn2021prototype}, where objects can have a \texttt{prototype} object, which acts as a template object that it inherits methods and properties from.
Tau Prolog defines three basic prototypes to represent Prolog objects:

\begin{itemize}
    \item The \texttt{pl.type.Var} prototype is used to represent logical variables in Prolog. The only argument that the constructor receives is the identifier of the variable as a string.
    \item The \texttt{pl.type.Num} prototype is used to represent numbers in Prolog. The constructor receives two arguments: the number representing the numeric value, and a boolean value which indicates if the value is a floating point number.
    \item The \texttt{pl.type.Term} prototype is used to represent atoms and compound terms in Prolog. The constructor receives a string identifying the term and, if the term is compound, an array of Prolog objects.
\end{itemize}

The \texttt{pl.type.Substitution} prototype is used to represent the substitutions in the resolution process and in the answers. The constructor receives, optionally, a JavaScript object binding variables with Prolog objects.

\begin{example}
The answer ``\texttt{X=foo(Z), Y=3}'' is represented by the following JavaScript object:
\begin{verbatim}
new pl.type.Substitution({
    "X": new pl.type.Term("foo", [new pl.type.Var("Z")]),
    "Y": new pl.type.Num(3, false)
});
\end{verbatim}
\end{example}

The \texttt{pl.type.Session} and \texttt{pl.type.Thread} prototypes are used to represent sessions and threads of execution, respectively. The user must create a session in order to load programs. When a session is created, a new inner thread is associated to the session by default. The threads forked from the same session share some information, such as the knowledge base (facts and rules) and the operators table, but other elements such as the choice point stack are thread-dependent. Here, threads are simply objects that contain information about derivations, allowing to query several goals simultaneously in the same session. As mentioned above, JavaScript is single threaded, so Tau Prolog threads have nothing to do with parallel execution of multiple goals. Each thread contains its own stack. When a goal is queried, an entry is added to the thread's stack. An entry pushed on the stack in the resolution is called a \emph{choice point}. A choice point is composed of a goal, a substitution, and a reference to its parent state, and it is represented in Tau Prolog by the prototype \texttt{pl.type.State}.

\subsection{Modules and packages}

A Tau Prolog's package is a JavaScript file that defines one or more modules. A module is characterized by a name, a set of predicates, and a set of visible predicates. Tau Prolog offers some modules to work with lists, manipulate the DOM, get statistics, interact with the operating system, format text, do random operations, etc., but users can build and distribute their own packages. All built-in predicates required by the ISO Prolog Standard \cite{ISO13211} are defined in the \texttt{system} module, which is always visible from any module. There is another special module, the \texttt{user} module, which forms the initial working space of the user.

Modules can be defined as Prolog files and imported through the \texttt{use\_module/1} directive, or they can be defined as Tau Prolog's packages, which offers the opportunity to define predicates in JavaScript to incorporate functionality not available in pure Prolog. The structure of a package is as follows:

\begin{verbatim}
var pl;
(function(pl) {
    var name = "";
    var predicates = function() { return {}; };
    var visible = [];
    if(typeof module !== "undefined") {
        module.exports = function(tau) {
            pl = tau;
            new pl.type.Module(name, predicates(), visible);
        };
    } else {
        new pl.type.Module(name, predicates(), visible);
    }
})(pl);
\end{verbatim}
The set of predicates is an object indexed by predicate indicator (name/arity). The set of visible predicates is an array containing the predicate indicators of visible predicates. There are two ways to define predicates in a package: as a list of Prolog clauses (which is a high level description of programs as terms), and as a JavaScript function that directly manipulates the choice point stack. To define a predicate as a list of Prolog clauses, we must use the internal representation of the objects introduced in Section \ref{sec:objects}.

\begin{example}
The \texttt{append/3} predicate from the \texttt{lists} module of the Tau Prolog's  lists package concatenates two lists (for readability, we omit the ``\texttt{pl.type.}'' text for constructors):
\begin{verbatim}
var predicates = function() {
   return {"append/3": [
      // append([], X, X).
      new Rule(
         new Term("append", [new Term("[]"),new Var("X"),new Var("X")]),
         null
      ),
      // append([H|T], X, [H|S]) :- append(T, X, S).
      new Rule(
         new Term("append", [
            new Term(".", [new Var("H"),new Var("T")]),
            new Var("X"),
            new Term(".", [new Var("H"),new Var("S")])]
         ),
         new Term("append", [new Var("T"),new Var("X"),new Var("S")])
      )
   ]};
};
\end{verbatim}
Manually writing a Prolog predicate in this way can be tedious and prone to errors, so Tau Prolog objects include a \texttt{compile} method (which takes no arguments) to automatically generate this code. Therefore, to obtain this representation, the user only has to consult the Prolog program in a session and print out the compiled code.
\end{example}

Sometimes it is necessary to resort to JavaScript to implement some functionality, either because it is not possible to do it directly in Prolog, or for efficiency reasons. Therefore, a predicate can also be defined as a JavaScript function that directly manipulates the choice point stack of a thread. These functions take as parameters: \emph{i)} the thread of execution; \emph{ii)} the current choice point; and \emph{iii)} the selected atom of the current goal, that is, the leftmost atom of the current goal.
A common scheme when implementing these functions is to get the arguments of the selected atom, check if there are errors (instantiation, type, domain, permission, etc.) and, if necessary, manipulate the choice point stack.

\begin{example}
The \texttt{random/3} predicate from the \texttt{random} module of the Tau Prolog's  random package generates a random number between two given numbers:
\begin{verbatim}
function(thread, point, atom) {
    var lower = atom.args[0], upper = atom.args[1], rand = atom.args[2];
    if(pl.type.is_variable(lower) || pl.type.is_variable(upper)) {
        thread.throw_error(pl.error.instantiation(atom.indicator));
    } else if(!pl.type.is_number(lower)) {
        thread.throw_error(pl.error.type("number",lower,atom.indicator));
    } else if(!pl.type.is_number(upper)) {
        thread.throw_error(pl.error.type("number",upper,atom.indicator));
    } else if(!pl.type.is_variable(rand) && !pl.type.is_number(rand)) {
        thread.throw_error(pl.error.type("number",rand,atom.indicator));
    } else {
        if(lower.value < upper.value) {
            var float = lower.is_float || upper.is_float;
            var gen = lower.value + Math.random() *
                (upper.value - lower.value);
            if(!float) gen = Math.floor(gen);
            var unif = new pl.type.Term("=",
                [rand, new pl.type.Num(gen, float)]);
            thread.prepend([new pl.type.State(
                point.goal.replace(unif),
                point.substitution,
                point 
            )]);
        }
    }
}
\end{verbatim}
In the first line of this function, the arguments of the selected atom are collected in the variables \texttt{lower}, \texttt{upper} and \texttt{rand}. The first two arguments are then verified to be non-variable. If one of them is a variable, an instantiation error is thrown. Similarly, the type of the arguments is checked below, and a type error is thrown if needed. If there are no errors, a random number is generated between the two given numbers and a new choice point is inserted, where the third argument is bound to the random value.
\end{example}

 It is important to note that functions which implement these predicates do not return any value. Returning a value that evaluates to \texttt{true} instead of \texttt{undefined} indicates to Tau Prolog that the function is asynchronous. For a predicate to be unsuccessful, it suffices not to push a new choice point on the stack.
 
 \begin{example}
 \label{example:async}
The \texttt{sleep/1} predicate from the \texttt{os} module of the Tau Prolog's operating system package sleeps the execution thread a specified number of milliseconds. For instance, the goal ``\texttt{?- sleep(1000), X=a.}'' gives the answer ``\texttt{X=a}'' after a second.

\begin{verbatim}
function(thread, point, atom) {
    var time = atom.args[0];
    if(pl.type.is_variable(time)) {
        thread.throw_error(pl.error.instantiation(thread.level));
    } else if(!pl.type.is_integer(time)) {
        thread.throw_error(pl.error.type("integer", time, thread.level));
    } else {
        setTimeout(function() {
            thread.success(point);
            thread.again();
        }, time.value);
        return true;
    }
}
\end{verbatim}
If there are no errors, the \texttt{sleep/1} predicate uses the \texttt{setTimeout} function to perform an action after a few seconds, and returns the value \texttt{true} (only in case the function is going to succeed). This tells Tau Prolog that an asynchronous task has been run, and that it should no longer apply resolution steps, as the predicate itself will resume the inference process at some point. In the case of \texttt{sleep/1}, after a few seconds, a new choice point will be pushed into the stack, and the \texttt{Thread.prototype.again} method will be invoked to resume the inference.
\end{example}

\subsection{Resolution and unification}

The \texttt{pl.type.Thread} prototype provides the \texttt{step} method to perform a resolution step. The first choice point of the stack is removed and the leftmost atom is selected from it. Tau Prolog provides clause indexing by first argument. The corresponding clauses are collected, and a new choice point is pushed onto the stack for each clause whose head  unifies with the atom. Hence, backtracking is automatically done by executing inference steps (when a predicate does not succeed, it does not push anything on the stack and the next choice point will correspond to a previous state).

\begin{verbatim}
Thread.prototype.step = function() {
    if(this.points.length === 0)
        return;
    var point = this.points.pop();
    var atom = point.goal.select();
    var definition_module = this.lookup_module(atom);
    var clauses = this.lookup_clauses(definition_module, atom);
    if(clauses instanceof Function) {
        return clauses(this, point, atom);
    } else {
        var states = [];
        for(var i in clauses) {
            var clause = clauses[i].rename(this);
            var occurs_check = this.get_flag("occurs_check");
            var mgu = pl.unify(atom, rule.head, occurs_check);
            if(mgu !== null) {
                var state = new State();
                state.goal = point.goal.replace(rule.body).apply(mgu);
                state.substitution = point.substitution.apply(mgu);
                state.parent = point;
                states.push(state);
            }
        }
        this.prepend(states);
    }
}
\end{verbatim}

Unification is implemented following the algorithm described by \cite{martelli1982efficient}. The \texttt{pl.unify} method takes two Prolog objects, and a flag to indicate whether the occurs check should be performed or not. User-defined Prolog term types can extend unification by implementing the \texttt{unify} method in its prototype (that takes the second Prolog object and the occurs check flag as arguments).
When two Prolog objects unify, the \texttt{unify} method returns a substitution. Otherwise, it returns \texttt{null}.

\begin{example}
Tau Prolog's JavaScript package defines a new term type, \texttt{pl.type.JSValue}, for manipulating JavaScript values that cannot be directly converted to Prolog. A \texttt{JSValue} unifies with other Prolog term if both are \texttt{JSValue} with the same value, or if one of them is a \texttt{JSValue} that contains a JavaScript object and the other one is a Prolog term of the form \texttt{\{prop\_1: value\_1, prop\_2: value\_2, ..., prop\_n: value\_n\}}, where \texttt{prop\_i} is an atom whose identifier represents a property of the JavaScript object, and \texttt{value\_i} is a Prolog term that unifies with the value of that property, for $1 \leq i \leq n$.

\begin{verbatim}
pl.type.JSValue.prototype.unify = function(obj, occurs_check) {
    if(pl.type.is_js_object(obj) && this.value === obj.value)
        return new pl.type.Substitution();
    if(pl.type.is_term(obj) && obj.indicator === "{}/1") {
        var left = [], right = [];
        var pointer = obj.args[0];
        var props = [];
        while(pl.type.is_term(pointer) && pointer.indicator === ",/2") {
            props.push(pointer.args[0]);
            pointer = pointer.args[1];
        }
        props.push(pointer);
        for(var i = 0; i < props.length; i++) {
            var bind = props[i];
            if(!pl.type.is_term(bind) || bind.indicator !== ":/2")
                return null;
            var name = bind.args[0];
            if(!pl.type.is_atom(name) || !this.value.hasOwnProperty(name.id))
                return null;
            var value = pl.fromJavaScript.apply(this.value[name.id]);
            right.push(bind.args[1]);
            left.push(value);
        }
        return pl.unify(left, right, occurs_check);
    }
    return null;
};
\end{verbatim}
Suppose there is an object ``\texttt{var o = \{x: 1, y: false, z: \{w: [2,"a"]\}\}};'' defined in the global scope of JavaScript. Then, the following goals are possible (the predicate \texttt{get\_prop/2} is described in detail in Section \ref{sec:web-interaction}):

\begin{verbatim}
?- get_prop(o, Object).
Object = javascript<object>.

?- get_prop(o, Object1), get_prop(o, Object2), Object1 = Object2.
Object1 = javascript<object>, Object2 = javascript<object>.

?- get_prop(o, {x: X, y: Y, z: {w: W}}).
X = 1, Y = false, W = [2,a].

?- get_prop(o, Object), Object = {x: X, y: false}, Object = {z: {w: W}}.
Object = javascript<object>, X = 1, W = [2,a].
\end{verbatim}
\end{example}

\subsection{Callback-based approach}

A \textit{callback function} is a function passed into another function as an argument, which is then invoked inside the outer function to complete some action. The \texttt{pl.type.Session} and \texttt{pl.type.Thread} prototypes have three main methods to load programs and query goals, all of them based on executing a user's callback when the main action has finished:

\begin{itemize}
    \item The \texttt{consult} method takes the input program and a callback and, if there are no syntax errors, it adds the parsed rules to the knowledge base, executing the callback afterwards. The \texttt{consult} method can take a string with the Prolog program, an URL or path to a Prolog file, or the identifier of a \texttt{<script type="text/prolog">} tag inserted on the webpage.
    \item The \texttt{query} method takes an input goal as a string and a callback and, if there are no syntax errors, it adds the goal to the stack of choice point and executes the callback. At first glance, consulting a program or querying a goal may not seem like an asynchronous task, but it is necessary to incorporate certain Prolog features such as term and goal expansion \cite{carlsson1988sicstus}, which require running Prolog code while parsing.
    \item The \texttt{answer} method takes a callback and pushes it to an internal stack of calls. The callback will be executed when the next computed answer is found. If there are no previous calls on the stack, the \texttt{again} method is invoked to start the search. 
\end{itemize}
After querying a goal, the search for a computed answer does not begin until the \texttt{answer} method is invoked.

\begin{verbatim}
Thread.prototype.answer = function(callback) {
    this.__calls.push(callback);
    if(this.__calls.length == 1)
        this.again();
};
\end{verbatim}
The \texttt{again} method performs resolution steps while there are choice point on the stack, as long as there are no errors and the resolution limit is not reached. When an answer is found, the user's callback is executed to handle it and the callback is removed from the stack. If the \texttt{step} method, that performs a resolution step, returns \texttt{true}, the search stops. This tells the thread that an asynchronous action has been taken, and the predicate will resume the search (by explicitly calling the \texttt{again} method) when it is done. This model allows predicates to perform asynchronous tasks, freeing up the browser until the search must continue. The Example \ref{example:async} shows the implementation of an asynchronous predicate that sleeps the thread a given amount of time.

\begin{verbatim}
Thread.prototype.again = function() {
    while(this.__calls.length > 0) {
        while(this.current_limit > 0
           && this.points.length > 0
           && this.head_point().goal !== null
           && !pl.type.is_error_state(this.head_point()))
                  if(this.step() === true)
                      return;
        var callback = this.__calls.shift();
        var answer = this.points.pop();
        (function(answer, callback) {
            setTimeout(function() {
                callback(answer);
            }, 0);
        })(answer, callback);
    }
};
\end{verbatim}

The following is a general scheme of how to use Tau Prolog to load a program and look for an answer. This outline can be cumbersome for the programmer, since up to three callbacks have to be nested to execute a goal. Furthermore, as can be seen, a callback or an object that contains callbacks can be passed for the different scenarios that may arise (success, failure, error or resolution limit).
\begin{verbatim}
const session = pl.create();
const program = "your prolog program";
const goal = "your prolog goal";
session.consult(program, {
    success: function() {
        session.query(goal, {
            success: function(goal) {
                session.answer({
                    success: function(answer) { /* Answer */ },
                    error:   function(err)    { /* Uncaught error */ },
                    fail:    function()       { /* Failure */ },
                    limit:   function()       { /* Limit exceeded */ }
                })
            },
            error: function(err) { /* Error parsing goal */ }
        });
    },
    error: function(err) { /* Error parsing program */ }
});
\end{verbatim}
As seen in \cite{kambona2013evaluation}, this problem is commonly known as the \textit{callback hell}. To overcome the drawbacks of this interface based on callbacks, Tau Prolog distributes a package that extends the prototypes to add a new interface based on promises, that we introduce in the next section.
However, notice that, unlike Tau Prolog, browsers can freeze while running goals that takes so long in a non-asynchronous Prolog implementation like jsProlog.

\subsection{Promises and reactive programming}

A promise object represents the eventual completion or failure of an asynchronous operation and its resulting value. Tau Prolog's promises package extends the \texttt{pl.type.Session} and \texttt{pl.type.Thread} prototypes to add new methods for consulting programs and querying goals, returning promises.

The Tau Prolog's promises package defines three new methods:

\begin{itemize}
    \item \texttt{promiseConsult}: consults a program and returns a promise that is resolved when the program loads successfully, or rejected when there is an error. It takes the same arguments as the \texttt{consult} method.
    \item \texttt{promiseQuery}: queries a goal and returns a promise that is resolved when the goal loads successfully, or rejected when there is an error. It takes the same arguments as the \texttt{query} method.
    \item \texttt{promiseAnswer}: finds the next computed answer and returns a promise that is resolved when it finds an answer or there are no more answers, or is rejected when there is an error or the limit of inferences has been reached. It takes the same arguments as the \texttt{answer} method.
\end{itemize}
Also, the package defines a fourth method, \texttt{promiseAsnwers}, to find all computed answers, returning an asynchronous generator.

\begin{example}\label{example:promise}
The following Node.js script loads a Prolog program that defines the \texttt{append/3} predicate, which concatenates two lists, and writes to the standard output all the computed answers for goal ``\texttt{?- append(X, Y, [a,b,c])}''. 
\begin{verbatim}
const pl = require("tau-prolog");
require("tau-prolog/modules/promises.js")(pl);

(async() => {
    const program = `
        append([], X, X).
        append([H|T], X, [H|S]) :- append(T, X, S).
    `;
    const goal = "append(X, Y, [a,b,c]).";
    const session = pl.create();
    await session.promiseConsult(program);
    await session.promiseQuery(goal);
    for await (let answer of session.promiseAnswers())
        console.log(session.format_answer(answer));
})();
\end{verbatim}
This script prints out the expected computed answers ``\texttt{X=[], Y=[a,b,c]}'', ``\texttt{X=[a], Y=[b,c]}'', ``\texttt{X=[a,b], Y=[c]}'' and ``\texttt{X=[a,b,c], Y=[]}''.
\end{example}

\cite{bainomugisha2013survey} define \textit{reactive programming} as a declarative programming paradigm that facilitates the development of event-driven and interactive applications by providing abstractions to express time-varying values and automatically managing dependencies between  such values.

It is possible to easily use the promises interface of Tau Prolog together with reactive programming libraries, as long as they support the conversion of asynchronous generator functions to \textit{observables}, such as RxJS \cite{clow2018observers} from version 7.0.0. An observable represents the idea of an invokable collection of future values or events.

\begin{example}\label{example:rxjs}
RxJS is a library for reactive programming using observables, to make it easier to compose asynchronous or callback-based code. The following Node.js script does the same result as the script in Example \ref{example:promise}, but creating an observable \texttt{ap} from the goal ``\texttt{?- append(X, Y, [a,b,c])}'' to which we subscribe to handle each of the answers.
\begin{verbatim}
const Rx = require("rxjs");
const pl = require("tau-prolog");
require("tau-prolog/modules/promises.js")(pl);

(async() => {
    const program = `
        append([], X, X).
        append([H|T], X, [H|S]) :- append(T, X, S).
    ';
    const goal = "append(X, Y, [a,b,c]).";
    const session = pl.create();
    await session.promiseConsult(program);
    await session.promiseQuery(goal);
    const ap = Rx.from(session.promiseAnswers());
    const sub = ap.subscribe(x => console.log(session.format_answer(x)));
})();
\end{verbatim}
Both, promises and observables, can take the place of answer handlers, but there are key differences. For example, observable subscriptions are cancellable, promises not: unsubscribing removes the listener from receiving further answers, and notifies the subscriber function to cancel work.
\end{example}

The JavaScript interface of SWI-Prolog pengines is based on HTTP requests and callbacks, and therefore, the same observable-based reactive interface can be built on top of it, just like Tau Prolog does, providing a generalized interface for Prolog systems.

\section{Web interaction with Tau Prolog}
\label{sec:web-interaction}

In this section, we focus on the main packages available on Tau Prolog for the web interaction: the DOM package and the Tau Prolog foreign function interface.

\subsection{DOM manipulation}

As defined by \cite{world2004document}, the \textit{Document Object Model} (DOM) is an API for accessing and manipulating HTML and XML documents, which are presented as node trees.
Tau Prolog's DOM package defines new term types and the \texttt{dom} module, which adds predicates that allow the user to modify the DOM:

\begin{itemize}
    \item Selector predicates provides methods that make it quick and easy to retrieve element nodes from the DOM by matching against properties. The \texttt{dom} module includes three non-deterministic predicates to look for DOM elements: \texttt{get\_by\_id/2}, \texttt{get\_by\_class/2} and \texttt{get\_by\_tag/2}. If there are no elements in the DOM with the specified identifier, class or tag, the predicates fail silently. If there is more than one, they find all of them on backtracking. 
    
    \item The \texttt{dom} module also includes predicates to go through the DOM starting at the HTML objects retrieved with the previous predicates: \texttt{parent\_of/2} and \texttt{sibling/2}.
    
    \item The \texttt{create/2} predicate takes an atom standing for an HTML tag and creates a new HTML object. Newly created HTML objects can be inserted in the DOM using the following predicates: \texttt{append\_child/2}, \texttt{insert\_after/2} and \texttt{insert\_before/2}. If we try to insert an element which is already part of the DOM, the predicate fails and the element is not inserted again. Otherwise, the element is inserted and the predicate succeeds.
    
    \item The following predicates allows the user to consult or modify the content, attributes and styles of HTML objects: \texttt{get\_attr/3}, \texttt{set\_attr/3}, \texttt{get\_html/2}, \texttt{set\_html/2}, \texttt{get\_style/3}, \texttt{set\_style/3}, \texttt{add\_class/2}, \texttt{remove\_class/2} and \texttt{has\_class/2}.
    
    \item Lastly, this module also includes predicates to create animations, such as \texttt{hide/1}, \texttt{show/1} or \texttt{toggle/1}, that hides, shows, or toggles the visibility of an HTML object, respectively.
\end{itemize}

\subsection{Event handling}

The \texttt{dom} module also enables the dynamic assignation of events, in order to run a Prolog goal when some browser event is triggered. The \texttt{bind/4} method takes an HTML object, an atom representing an event type (\texttt{click}, \texttt{mouseover}, \texttt{mouseout}, etc.), an event and a goal, and bind the HTML object with said goal for that type of event. The third argument is bound to a new term that represents an HTML event, from which we can read information using the \texttt{event\_property/3} predicate.

\begin{example}
In this example, the \texttt{keypress} event has been added to the page body, so when a key is pressed, the HTML object whose id is \texttt{output} displays what key has been pressed.
\begin{verbatim}
get_by_id(output, Output),
get_by_tag(body, B),
bind(B, keypress, Event, (
        event_property(Event, key, Key),     
        set_html(Output, Key)      
    )
).
\end{verbatim}
Notice that the \texttt{event\_property/3} predicate and the \texttt{Event} term only make sense inside an event's goal, since they don't hold any information until the event is triggered. Any time an event is triggered, Tau Prolog forks the session that assigned the event, and it runs the goal in the new thread (just for the first answer).
\end{example}

The \texttt{unbind/2} and \texttt{unbind/3} predicates allow us to remove the events attached to an HTML object. The  \texttt{prevent\_default/1} predicate allows us to prevent the browser default behaviour regarding an event (for instance, to keep a form from being sent). A list with the events supported by the browser can be found in \cite{mdn2021event}.
   
\subsection{Foreign function interface}

A foreign function interface is a mechanism allowing a program written in a programming language to call routines written in another. The Tau Prolog's JavaScript package defines the \texttt{js} module, whose predicates allows the user to invoke JavaScript functions from Prolog programs, send ajax requests, and manipulate JSON data.

The \texttt{js} module exports a few main predicates to invoke JavaScript functions:

\begin{itemize}
    \item \texttt{apply/4}: invokes a JavaScript function with a list of arguments. \texttt{apply(Context, Method, Arguments, Value)} is true if \texttt{Value} unifies with the result of calling the method \texttt{Method} of the JavaScript object \texttt{Context} with arguments \texttt{Arguments}. If \texttt{Method} is a JavaScript function, \texttt{Value} is the result of calling the JavaScript function in the context of the JavaScript object \texttt{Context}.
    \item \texttt{get\_prop/3}: gets a property of a JavaScript object. \texttt{get\_prop(Context, Property, Value)} is true if \texttt{Value} unifies with the value of the property \texttt{Property} of the JavaScript object \texttt{Context}.
    \item \texttt{global/1}: gets the global JavaScript object. \texttt{gobal(Context)} is true if and only if \texttt{Context} is the global JavaScript object (\texttt{window} in browser or \texttt{global} in Node.js).
\end{itemize}
There are also versions of \texttt{get\_prop/2} and \texttt{apply/3} predicates where the context is just the global object.

\begin{example}
The \texttt{Array.prototype.concat} method is used to merge two or more arrays, so it can be applied in the context of an array to get the concatenation of a list of lists. Similarly, the \texttt{String.prototype.concat} method can be applied in the context of a string to get the concatenation of a list of atoms:
\begin{verbatim}
?- apply([], concat, [[1,2],[3,4,5],[6]], Xs).
Xs = [1,2,3,4,5,6].

?- apply(`', concat, [hello, `, ', world, !], Str).
Str = `hello, world!'.
\end{verbatim}
Here, the first and third arguments are translated to JavaScript objects. Lists are converted into arrays, and atoms into strings. Then, the method is applied in the context of the first object, using the objects of the third argument as parameters. Finally, the result is translated back from JavaScript to Prolog.
\end{example}

Note that when a value can not be directly converted from JavaScript to Prolog, such as an object or a function, it is returned wrapped in a new type of term, \texttt{pl.type.JSValue}. JavaScript's objects can be explicitly converted from and to Prolog terms using the \texttt{json\_prolog/2} predicate.
%
%
The JavaScript foreign function interface of Tau Prolog is really simple but very effective. As we will see in the following section, it allows to create interfaces with complex JavaScript APIs, such as drawing on \texttt{<canvas>} HTML elements.

\section{Tau Prolog programming environment}
\label{sec:tools}

Although the main purpose of Tau Prolog is to embed Prolog code in web pages, Tau Prolog also has an online interactive interpreter which is very convenient for debugging, testing and sharing Prolog code. In this section, we show the programming environment for Tau Prolog.

\subsection{Tau Prolog Sandbox}

\begin{figure}
    \centering
    \includegraphics[width=\textwidth]{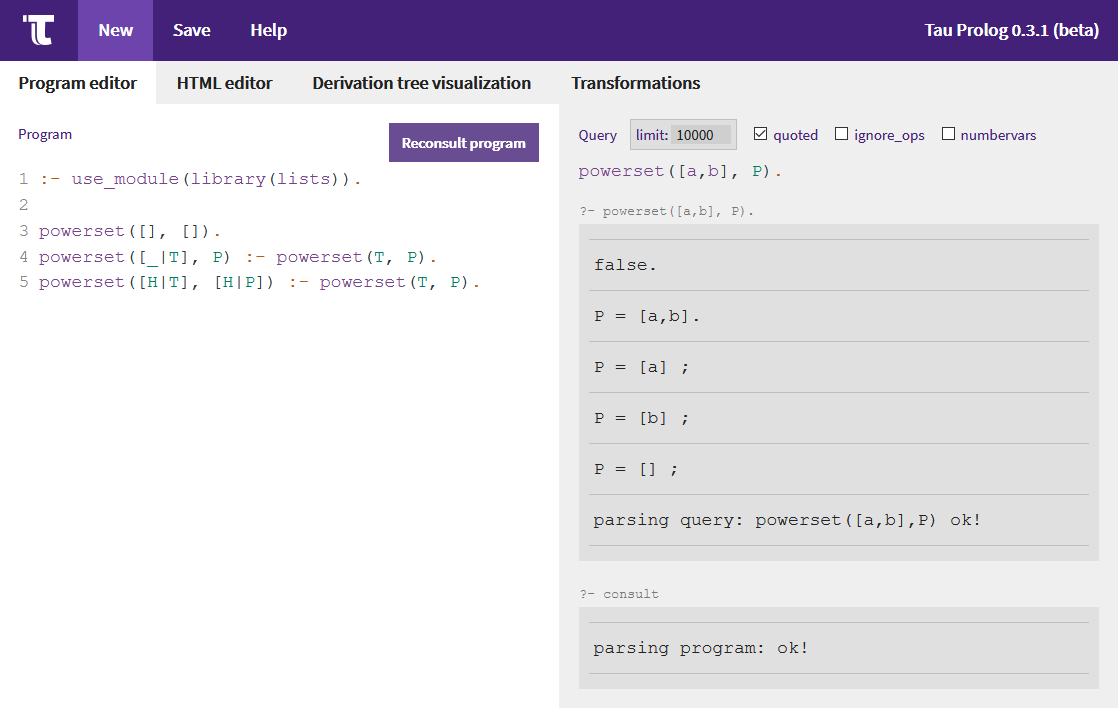}
    \caption{Screenshot of the Tau Prolog Sandbox online tool running a Prolog program.}
    \label{fig:sandbox}
\end{figure}

 The \textit{Tau Prolog Sandbox}\footnote{http://tau-prolog.org/sandbox} (see Figure \ref{fig:sandbox}) is an online interactive interpreter of Prolog which runs the latest version of all Tau Prolog's packages availables so far. Programs can be freely saved and shared via URL. The browser version of Tau Prolog has a virtual file system, so even programs including operating system interactions --such as manipulate files and directories-- can be executed in the sandbox without any problem (of course, interactions with the operating system in Node.js are real).
 
 Below the \textsf{Program editor} option, users can consult Prolog programs and query goals. In the \textsf{query} area, the user can set the limit of inferences in a derivation and the writing options defined by the ISO Prolog Standard \cite{ISO13211}.
 After consulting a Prolog program, under the \textsf{HTML editor} choice users can insert HTML code and run programs related to it.
 
\begin{figure}
    \centering
    \includegraphics[width=\textwidth]{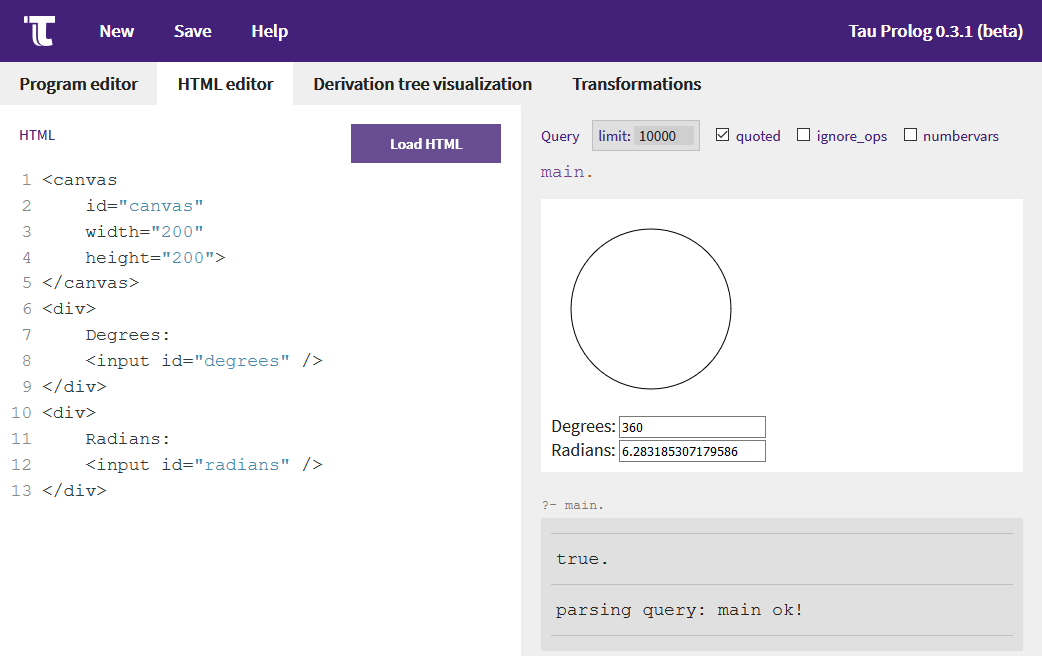}
    \caption{Screenshot of the Tau Prolog Sandbox online tool under the HTML editor.}
    \label{fig:html}
\end{figure}

 \begin{example}
 The Canvas API provides a means for drawing graphics via JavaScript and the HTML \texttt{<canvas>} element.
 The following Prolog program creates an animation in a \texttt{<canvas>} element. The \texttt{draw/2} predicate takes the context of the canvas and the degrees, and draws an arc on it. The \texttt{main/0} predicate draws arcs from 0 to 360 degrees every 10 milliseconds.
 
 \begin{verbatim}
:- use_module(library(os)).
:- use_module(library(js)).
:- use_module(library(dom)).

draw(Ctx, Degrees) :-
    apply(Ctx, clearRect, [0, 0, 200, 200], _),
    Radians is Degrees * pi / 180,
    apply(Ctx, beginPath, [], _),
    apply(Ctx, arc, [100, 100, 80, 0, Radians], _),
    apply(Ctx, stroke, [], _),
    get_by_id(radians, Rad), set_attr(Rad, value, Radians),
    get_by_id(degrees, Deg), set_attr(Deg, value, Degrees).

main :-
    get_by_id(canvas, Canvas),
    apply(Canvas, getContext, [`2d'], Ctx),
    between(0, 360, Degrees),
    draw(Ctx, Degrees),
    sleep(10),
    false ; true.
\end{verbatim}
Figure \ref{fig:html} shows the \textsf{HTML editor} tab after running the animation. In this example, we are using the JavaScript foreign function interface for drawing the canvas, and the DOM package to get the HTML elements and write into the \texttt{<input>} elements.
 \end{example}

The \textsf{Transformations} tab lists all the clauses loaded into the session, and allows the user to apply some automatic transformations to the code, such as the unfolding transformation defined in \cite{tamaki1984unfold}. This tab is especially convenient for viewing code transformations that a Prolog interpreter applies (body conversion, term and goal expansion, Definite Clause Grammar notation, etc.).

Finally, the \textsf{Derivation tree visualization} tab draws the derivation tree for Prolog goals. This tool is not exclusive of the Tau Prolog Sandbox, in fact, it is an extension of Tau Prolog that users can install in their own web pages, so we introduce it in the following section.

\subsection{Graphical derivation trees}

The graphical derivation trees\footnote{\url{https://github.com/tau-prolog/draw-derivation-trees}} tool extends the \texttt{pl.type.Session} and \texttt{pl.type.Thread} prototypes adding a new method, \texttt{draw}, that takes the following arguments:

\begin{itemize}
    \item the maximum number of answers to find in the derivation (to avoid infinite trees),
    \item the \texttt{<canvas>} HTML element, or its identifier,
    \item and (optionally) a JavaScript object with style properties.
\end{itemize}
The \texttt{draw} method must be called after querying a goal instead of the \texttt{answer} method.

\begin{figure}
    \centering
    \includegraphics[width=\textwidth]{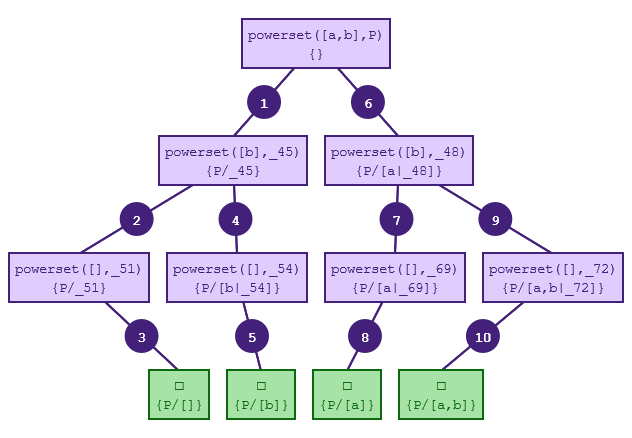}
    \caption{Graphical derivation tree generated by Tau Prolog.}
    \label{fig:tree}
\end{figure}

\begin{example}
The following HTML document draws the derivation tree for goal ``\texttt{?- powerset([a,b], P)}'', where \texttt{powerset/2} generates all the subsets of a given set:
\begin{verbatim}
<html>
  <head>
    <script src="tau-prolog/core.js"></script>
    <script src="tau-prolog/draw-derivation-trees.js"></script>
    <script id="program.pl" type="text/prolog">
        powerset([], []).
        powerset([_|T], P) :- powerset(T, P).
        powerset([H|T], [H|P]) :- powerset(T, P).
    </script>
  </head>
  <body>
    <canvas id="derivation"></canvas>
    <script type="text/javascript">
        var session = pl.create();
        session.consult("program.pl", function() {
            session.query("powerset([a,b], P).", function() {
                session.draw(10, "derivation");
            });
        });
    </script>
  </body>
</html>
\end{verbatim}
Figure \ref{fig:tree} shows the graphical derivation tree generated by Tau Prolog for this goal. Each node displays the current goal and substitution in the inference process, where answers are leaves with empty goals ``$\square$''. Note that only variables in the original goal are shown in substitutions. Nodes representing answers and errors are highlighted with different colors.
\end{example}

\section{Benchmarks}
\label{sec:benchmarks}

{
\color{black}

In this section, we present some benchmarks to compare the performance of Tau Prolog with other Prolog systems.
Table \ref{tab:runtime_browsers} shows the average runtime of executing Prolog benchmarks with Tau Prolog 0.3.4 running in Chromium 109.0.5414.119, SWI-Prolog 8.4.2, the latest version\footnote{\url{https://github.com/mthom/scryer-prolog/tree/1118b37c9232d10b7c2c84648bf364d859210319}} of Scryer Prolog, the latest version\footnote{\url{https://github.com/trealla-prolog/trealla/tree/c213efefe886c830d9cc98de2ff7e9ee6d2d28b7}} of Trealla Prolog, GNU Prolog 1.4.5, Ciao Prolog 1.22.0, the latest version\footnote{\url{https://www.xlog.ch/izytab/doclet/docs/18_live/10_reference/example02/package.html}} of Dogelog running in Chromium 109.0.5414.119, the latest version\footnote{\url{https://github.com/CapelliC/hhprolog/tree/87074de0aea959ed0ac2c0eaa45a6aa13e3e37d5}} of hitchhiker Prolog running in Chromium 109.0.5414.119 and the latest version\footnote{\url{https://github.com/cubiwan/jsProlog/tree/5d6f00e7d5cc435a1bf706527e7c4781957170cd}} of jsProlog running in Node 18 (since it has been necessary to increase the memory limit for some benchmarks). All tests have been executed in Ubuntu 22.04.1 LTS (64 bits) using a desktop computer equipped with an Intel\textregistered\ Core\texttrademark\ i7-8700 CPU @ 3.20 GHz and 32.00 GB RAM. Test programs considered for benchmarking can be found at \url{https://github.com/tau-prolog/tau-prolog/tree/master/examples/tplp}, and all goals were chosen to give a reasonably long overall time. For instance, the queens benchmark refers to finding all solutions to the 8 queens problem.
Currently, we can see that classic Prolog systems are much faster than Tau Prolog. However, Tau Prolog is capable of performing common tasks in the browser (validating forms, creating animations,\footnote{\url{http://tau-prolog.org/examples/draggable}} deploying simple games,\footnote{\url{http://tau-prolog.org/examples/snake}} etc.) without any problem, as we show in \cite{riaza2021tau}.
Blank entries in the hitchhiker Prolog column (*) are due to the lack of extra-logical predicates (control constructs, arithmetic operators, etc.). Blank entries in the jsProlog column (**) correspond to errors due to memory limits.

\begin{table}[!tb]
\centering

\color{black}

\caption{Average runtime (in seconds) of Prolog programs on different systems.}
\label{tab:runtime_browsers}
\begin{tabular}{llllllllll}
\hline
Benchmark  & Tau & SWI & Scryer & Trealla & GNU & Ciao & Dogelog & hh & jsProlog \\
\hline
mergesort  & 0.785 & 0.010 & 0.006 & 0.008 & 0.004 & 0.001 & 0.016 & *     & 160.352 \\
zebra      & 0.920 & 0.009 & 0.027 & 0.032 & 0.005 & 0.004 & 0.019 & 0.113 & 62.851 \\
queens     & 1.213 & 0.037 & 0.021 & 0.020 & 0.014 & 0.005 & 0.075 & *     & 228.594 \\
peano      & 4.208 & 0.005 & 0.003 & 0.006 & 0.004 & 0.000 & 0.014 & 0.061 & ** \\
sat        & 4.517 & 0.016 & 0.062 & 0.069 & 0.060 & 0.026 & 0.171 & *     & ** \\
inorder    & 5.048 & 0.003 & 0.004 & 0.008 & 0.003 & 0.000 & 0.009 & 0.025 & ** \\
append     & 6.284 & 0.001 & 0.002 & 0.002 & 0.001 & 0.000 & 0.005 & 0.135 & ** \\
\hline
\end{tabular}
\end{table}

}

\section{Conclusions and future work}

In this paper we have described the features and implementation details of Tau Prolog, a client-side Prolog interpreter for the Web fully implemented in JavaScript, ISO Prolog Standard compliant, which allows to execute --possibly asynchronous-- Prolog programs in the browser overcoming the limitations of the JavaScript concurrency model. We have shown the main Tau Prolog's packages to interact with the Web, including a foreign function interface to invoke JavaScript code from Prolog, and the Tau Prolog's programming environment. These packages reflect the main advantages of our proposal compared to other server-side Prolog interpreters. Finally, we have shown some benchmarks to measure the performance of Tau Prolog. As future work, we hope to improve the performance and efficiency of Tau Prolog (by implementing, e.g., last call optimization), and to incorporate features of Constraint Handling Rules \cite{fruhwirth1998theory} and Constraint Logic Programming \cite{jaffar1987constraint} to it. In this line, we consider implementing an interface for attributed variables, originally defined by \cite{holzbaur1990}.

\section{Acknowledgments}

We thank Miguel Riaza for implementing the parser of Tau Prolog, and for his help in the core development. We thank Dr. Jose Maria Garcia-Garcia for documenting most of the predicates availables on Tau Prolog until the present. We thank Dr. Paulo Moura for his help testing Tau Prolog and fixing many bugs of built-in predicates, and for his work on integrating Tau Prolog with Logtalk. We thank Dr. Markus Triska for providing his module for formatting text, originally written for Scryer Prolog. We thank Jan Burse for his help in reporting bugs. We thank Dr. Ginés Moreno and Dr. Pascual Julián-Iranzo for comments on the manuscript.

\bibliographystyle{acmtrans}
\bibliography{taupl}

\end{document}